\documentclass[twocolumn,showpacs,preprintnumbers,prd,nofootinbib]{revtex4-1}
\usepackage{latexsym}
\usepackage{graphicx}

\begin{document}
\title{\bf 2D Radial Distribution  Function of Silicene}

\author{M. R. Ch\'avez-Castillo$^{1,2}$, M. A. Rodr\'\i guez-Meza$^2$, and L. Meza-Montes$^1$} 
\affiliation{
$^1$Instituto de F\'isica, Benem\'erita Universidad Aut\'onoma de Puebla 
 Apdo. Postal J-48, Puebla, Pue. 72570 M\'exico.  \\ 
$^2$Instituto Nacional de Investigaciones Nucleares 
 Apdo. Postal 18-1027, M\'exico, D.F. 11801 M\'exico. \\ 
 e-mails: mchavez@ifuap.buap.mx; 
 marioalberto.rodriguez@inin.gob.mx; lilia@ifuap.buap.mx
 }

\begin{abstract}
\noindent
Silicene is the counterpart of graphene and its potential applications as a part
of the current electronics, based in silicon, make it a very important system to study. We perform molecular dynamics simulations and analyze the structure of a two dimensional array of Si atoms by means of the radial distribution function, at different temperatures and densities. As a first approach, the 2D Lennard-Jones potential is used and two sets of parameters are tested. We find that the radial distribution function does not change with the parameters and resembles the corresponding to the (111) surface of the FCC structure. The liquid phase appears at very high temperatures, suggesting a very stable system in the solid phase.

\medskip
\noindent
\textit{Keywords}: 
 Silicene; Radial distribution function; Molecular-dynamics

\bigskip
\noindent
El siliceno es la contraparte del grafeno y sus potenciales aplicaciones en la electr\'onica actual, basada en el silicio, lo hacen un objeto de estudio muy importante. Realizamos simulaciones de din\'amica molecular de un sistema bidimensional formado por \'atomos de Si, a diferentes temperaturas y densidades, analizando la estructura por medio de la funci\'on de distribuci\'on radial. Como primera aproximaci\'on, usamos el potencial de Lennard-Jones en 2D. Se encontr\'o que utilizando dos conjuntos de par\'ametros diferentes, la funci\'on de distribuci\'on radial no cambia y se asemeja a la del plano (111) de la estructura FCC. La fase l\'iquida aparece a muy altas temperaturas, sugiriendo un sistema muy estable en la fase s\'olida.

\medskip
\noindent
\textit{Descriptores}: 
Siliceno; Funci\'on de distribuci\'on radial; Din\'amica molecular

\end{abstract}

\date{ \today}
\pacs{61.46.-w; 61.48.-c;68.65.-k}
\preprint{}
\maketitle

\section{Introduction}

During the last two decades, honeycomb-structured materials became very important in nanoscience.
Their unique orbital symmetry gives rise to exceptional properties in these quasi-1D carbon-based systems, such as fullerenes, nanotubes, graphene and its nanoribbons \cite{Kroto-1985, Iijima-1993, Novoselov-2004}. Among them, graphene is doubtless the most studied at present time, due to its potential applications and the fact that is the first bidimensional material of one-atom thickness \cite{Castro-2009, DasSarma-2011, Aufray-2010}.
On the other hand, the interest on fabrication, characterization and study of silicon nanostructures keeps continously expanding. The possibility of having a graphene-like structure, the silicene, has been studied theoretically  already for several years \cite {Cahangirov-2009, Houssa-2010, Yu-2010, Takeda-1994}. From an electronic point of view, silicene could be equivalent to graphene, having the advantage of an easy integration to the present electronics, which is based on bulk silicon. Moreover, its borders do not react with oxygen \cite{De Padova-2008}. Nevertheless, it has a drawback. Less versatile than carbon, silicon hardly hibridizes to $sp ^2$ bonds. Therefore, syntesis and growth of silicene is extremely hard \cite{Lebegue-2009}.
Only recently, silicon structures resembling graphene, such as self-assembled silicene nanoribbons \cite{Aufray-2010} and silicene sheets deposited on silver crystals \cite{Lalmi-2010} have been reported. Growth on a ZrB$_2$  substrate has been succesful \cite{Fleurence-2011}.
For the hexagonal lattice, the lattice constant $a$ for Si and Ge, as reported by Leb\'egue and Eriksson \cite{Lebegue-2009} are 3.860 \AA\ and 4.034 \AA\  respectively, which are larger than the one for graphene (2.46 \AA) since these atoms have larger radii.
Cahangirov \emph{et al.}  \cite{Cahangirov-2009} showed that the bonding distance Si-Si (Ge-Ge) is 2.25 \AA (2.38 \AA).

With four \emph{sp} electrons  in valence band, Si and Ge are chemically very similar to C, in spite of this they behave very differently. The difference in the chemistry exhibited by carbon and silicon can be traced to the difference in their $\pi$ bonding capabilities. First, the energy difference between the valence \emph{s} and \emph {p} orbitals for carbon is about twice that for silicon (Si: E$_{3p}$-E$_{3s}$ = 5.66 eV, C: E$_{2p}$-E$_{2s}$=10.60 eV) \cite{Boon-2007}); as a result, silicon tends to utilize all three of its valence \emph{p} orbitals, resulting in \emph{sp}$^3$ hybridization while, in contrast the relatively large hybridization energy for carbon implies that this will "activate" one valence \emph{p} orbital at a time, as requirements the bonding situation, giving rise, in turn, to \emph{sp}, \emph{sp}$^2$, and \emph{sp}$^3$ hybridizations. Second, the $\pi$-$\pi$ overlap decreases by roughly an order of magnitude in going from carbon to silicon due to the significant increase in atomic distance, resulting in much weaker $\pi$ bonding for silicon in comparison with that for carbon. Hence, Si=Si are in general much weaker than C=C bonds. \cite{Zhang-2002}. Fagan {\it et al.} \cite{Fagan-2001} establish the theoretical similarities and differences between Si and C nanotubes. For the silicon nanotubes (SiNTs) studied, they obtained a cohesive energy value of  0.83 eV/atom, which is higher than the total energy per atom for the diamond-like structure. Considering that the cohesive energy for the Si bulk in diamond structure is 4.63 eV/atom, the energies for the studied nanotubes are only 82\% of the bulk. Comparing with carbon nanotubes (CNTs) that have around of 99\% of the cohesive energy they would have in perfect crystalline graphite, these results help to understand the diffilculty to produce the SiNTs and, especially, silicene. Zhang {\it et al.} \cite{Zhang-2002} also studied silicon nanotubes and their results suggest that silicon nanotubes can in principle be formed, however, the energy-minimized SiNTs adopt a severely puckered structure (with a corrugated surface) with Si-Si distances ranging from 1.85 to 2.25 \AA.

Some silicene properties have been predicted theoretically to be very similar to their corresponding in graphene \cite{Lalmi-2010, Aufray-2010}.  Ab-initio calculations revealed that silicene clusters can be used in Field-Effect Transistors and hydrogene storage, and their electronic properties have been studied using molecular dynamics \cite{Jose-2011}. Ince and Erkoc (IE) determined the structure of silicene nanoribbons  of different widths and  lenghts \cite{Ince-2011}.

In this paper we report results on structural properties of a bidimensional array of Si atoms. By means of molecular dynamics, we determine the radial distribution function for different temperatures and densities. We analyze the effect of the two-body potential considering a Lennard-Jones potential with two sets of parameters, the IE parameters and those suggested by Stillinger and Weber (SW) for bulk Si \cite{Stillinger-1985}. Our calculations represent a description of  free-standing silicene, since ab-initio studies have shown that two dimensional honeycombe structures for Si and Ge are stable \cite{Cahangirov-2009}. Our model provides insight to understanding the structural properties of silicene. Taking into account potentials developed for silicon or interactions with the substrate for the case of adsorbed nanosheets will provide an improved description. In Section 2 we briefly describe our calculation method and the used parameters. In Section 3 the results are analized and finally, Conclusions close this report.

\section{Methodology}

We perform molecular 2D dynamics simulations. Si atoms interact via Lennard-Jones (LJ) potential
\begin{equation}
\label{LJ}
U(r)=4\varepsilon\left[\left( \frac{\sigma}{r}\right)^{12}- \left(\frac{\sigma}{r}\right)^6\right] \, .
\end{equation}

The first term is a positive (repulsive) short-range interaction, related to the
electrostatic repulsion. The second term, a negative contribution, is the Van der Waals potential. The parameter $\varepsilon$ ($\sigma$) corresponds to the energy (spatial range) scale.
LJ potential works well for rare gases and poorly for materials where many-body effects are important, like metals or semiconductors. Nevertheless, it can give a rough description and has been used as the two-body part in Si nanostructures \cite{Ince-2011, Galashev-2007}.
Details of our approach can be found in \cite{nagbody, Rodriguez-2011}.

We compare calculations performed with two sets of parameters shown in Table \ref{parameters}.
The first set takes $\sigma$ as the distance in the Si-Si dimer \cite{Ince-2011} while SW determined the values for  crystalline Si at 0 K \cite{Stillinger-1985}. They are alike, with the IE well being deeper and wider.

\begin{table}[h]
\centering
\scalebox{0.9}[0.9]{
\begin{tabular}{|c|c|c|c|c|c|}\hline 
$\sigma_{IE}$ &  $\varepsilon_{IE}$ \\ \hline
 0.2295 nm & 2.817 eV \\ 
 & (4.5134$ \times 10^{-19}$ J)   \\ \hline\hline
$\sigma_{SW}$ & $\varepsilon_{SW}$  \\ \hline
 0.20951 nm & 2.168 eV \\ 
& (3.4738$ \times 10^{-19}$ J)  \\ \hline
\end{tabular}}
\caption{LJ potential parameters, IE \cite{Ince-2011} and  SW \cite{Stillinger-1985}.}
\label{parameters}
\end{table}

We focus on the radial distribution function (RDF), which describes how the atoms in a system radially arrange around each other. The RDF gives information about the average structure of disordered molecular systems such as liquids. It is also of fundamental importance in thermodynamics, because some macroscopic thermodynamic quantities can be calculated using the RDF, for example the pressure and the energy. The RDF is a function of the radial distance \emph{r}, defined as

\begin{equation}
g(r)=\frac{2V}{N^2}\left<\sum_{i<j}{\delta(r-r_{ij})}\right> \, ,
\label{fdr}
\end{equation}
where \emph{V} is the volume, \emph{N} is the number of atoms, $r_{ij}$ is
the position vector of atom \emph{j} respect to the \emph{i}th atom, and the brackets indicate average over all atoms. This function $g(r)$
 has characteristic shapes for different phases \cite{Rodriguez-2011, Rapaport-1995}.

\section{Results}

Simulations with 512 atoms at different coverages (densities) and temperatures were performed.
Densities were taken in dimensionless units $\rho^*_{SW}$ ($\rho^*_{IE}$)= 0.459 (0.604) -diamond Si-,  0.592 (0.778) and 0.690 (0.907), the latter ones higher than liquid silicon density \cite{Ioffe}.
The chosen temperatures, T$^*_{SW}$ (T$^*_{IE}$)= 0.0119 (0.0092), 0.0670 (0.0515) and 0.1987 (0.1530), correspond to normal conditions, bulk melting point and a higher temperature, respectively, since the melting point of 2D structures is known to increase considerably \cite{Zakharchenko-2011}. Table \ref{realunits} shows the values of density and temperature in real units.
LJ potential tends to generate a FCC structure, and its cleavage plane is (111) \cite{Bechstedt-2003}. Besides, LJ tends to generate this kind of lattice. Thus, structures related to FCC are expected.

\begin{table}[h]
\centering
\scalebox{0.9}[0.9]{
\begin{tabular}{|c|c|c|c|c|c|c|c|c|c|c|c|}\hline 
\multicolumn{4}{|c|}{Density} &&  \multicolumn{4}{|c|}{Temperature} \\ \hline 
\multicolumn{2}{|c|}{$\rho$, gr/cm$^3$} & \multicolumn{2}{|c|}{$\rho^*_{SW}$ ($\rho^*_{IE}$)} && \multicolumn{2}{|c|}{T, K} & \multicolumn{2}{|c|}{$T^*_{SW}$ ($T^*_{IE}$)} \\ \hline
\multicolumn{2}{|c|}{2.33} & \multicolumn{2}{|c|}{0.459 (0.604)} && \multicolumn{2}{|c|}{300} & \multicolumn{2}{|c|}{0.0119 (0.0092)} \\ \hline 
\multicolumn{2}{|c|}{3.00} & \multicolumn{2}{|c|}{0.592 (0.778)}&& \multicolumn{2}{|c|}{1685} & \multicolumn{2}{|c|}{0.0670 (0.0515)} \\ \hline
\multicolumn{2}{|c|}{3.50} & \multicolumn{2}{|c|}{0.690 (0.907)}&& \multicolumn{2}{|c|}{5000} & \multicolumn{2}{|c|}{0.1987 (0.1530)}
 \\ \hline
\end{tabular}}
\caption{Density and temperature, in real units, for potential parameters IE \cite{Ince-2011} and  SW \cite{Stillinger-1985}.}
\label{realunits}
\end{table}

\begin{figure}[!htpb]
\centering
\includegraphics[width=6cm,height=9.6cm]{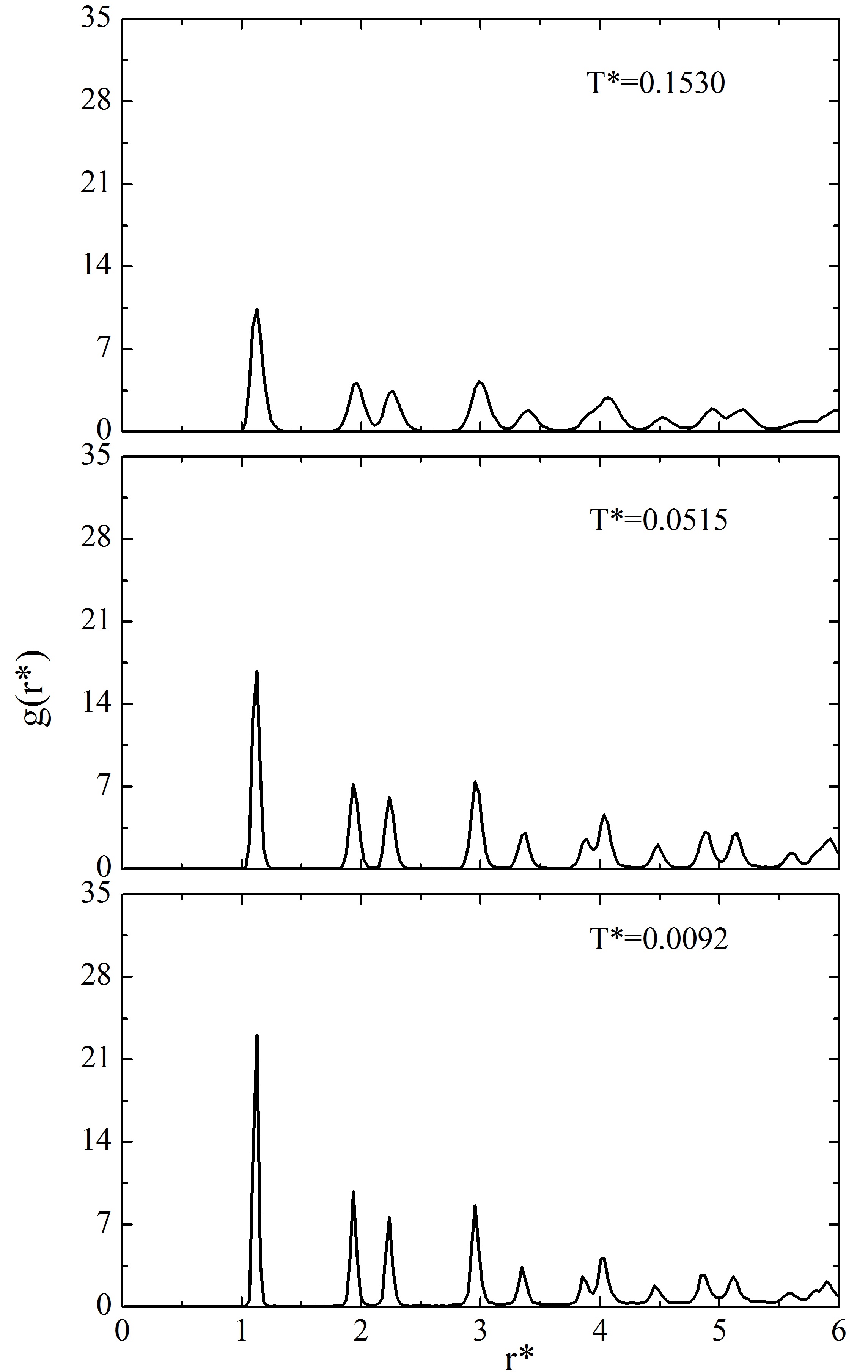}
\caption{Radial Distribution Function, IE parameters and  $\rho$*= 0.604 which corresponds
to crystalline (diamond) silicon density.}
\label{figrdfie233}
\end{figure}

\begin{figure}[!htpb]
\centering
\includegraphics[width=6cm,height=9.6cm]{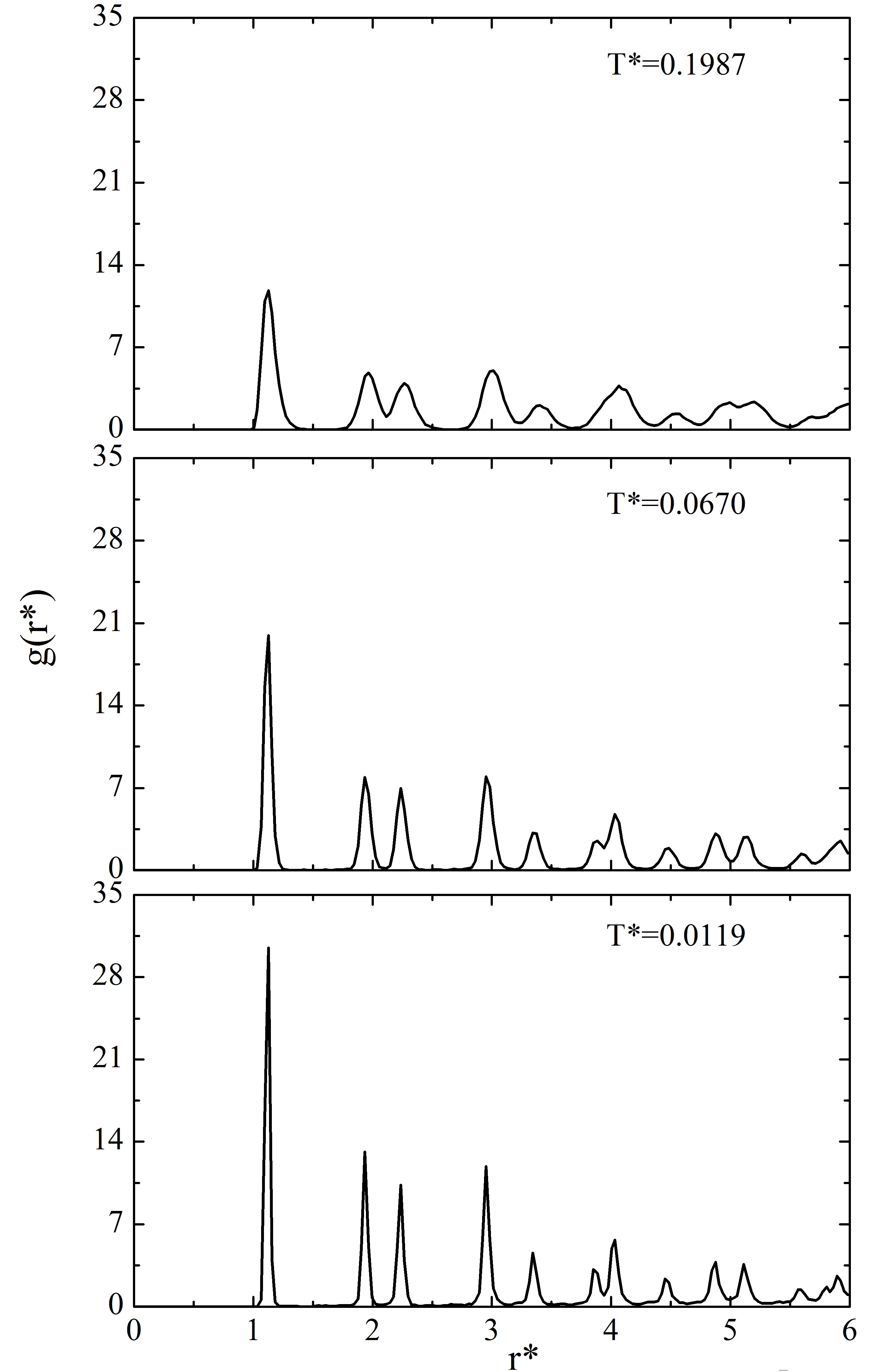}
\caption{Radial Distribution Function, SW parameters and  $\rho$*= 0.459, corresponding to crystalline (diamond) silicon density.}
\label{figrdfsw233}
\end{figure}

Figs. \ref{figrdfie233} and \ref{figrdfsw233} show the RDF's for IE and SW parameters. Here $r \ast= r/\sigma$. At  $\rho^*_{IE}$ ($\rho^*_{SW}$)= 0.604 (0.459) and T$^*_{IE}$ (T$^*_{SW}$)= 0.0092 (0.0119) peaks are narrow, a typical characteristic of the crystalline structure for solids. These RDF's coincide with the RDF in 2D obtained by Rodr\'iguez {\it et al.} \cite{Rodriguez-2011} for  Argon, but with a slight shift to the right. We see that even at very high temperatures the peaks slightly broaden while the positions of the peaks and the overall shape do not change, indicating a very high stability of the system. The RDF at T$^*_{IE}$ (T$^*_{SW}$)= 0.1530 (0.1987) resembles the radial distribution function for atoms in layer 1 of the (111) surface obtained by Broughton and Gilmer \cite{Broughton-1983}, when they  studied the structures of atomic layers in the crystal-vapor interfaces for FCC structures by molecular dynamics. 
This FCC surface is characteristic for 2D honeycomb structures, with the difference that it has an extra atom at the center of each hexagon. It is worth mentioning that compared to graphene, the FDR's obtained are similar, showing a difference in the position of the peaks \cite{Fasolino-2007, Zhao-2009}, with a scaling which depends on the value of the bond length, and increased peak heights due to contributions of the central atoms.

In the RDF at T$^*_{IE}$ (T$^*_{SW}$)= 0.1530 (0.1987) the peak heights diminish, because the short-range interaction between atoms is being lost and the system approaches to the liquid phase.
Similar results are found for the intermediate value of density, not shown here.

\begin{figure}[htpb]
\centering
\includegraphics[width=6cm,height=9.6cm]{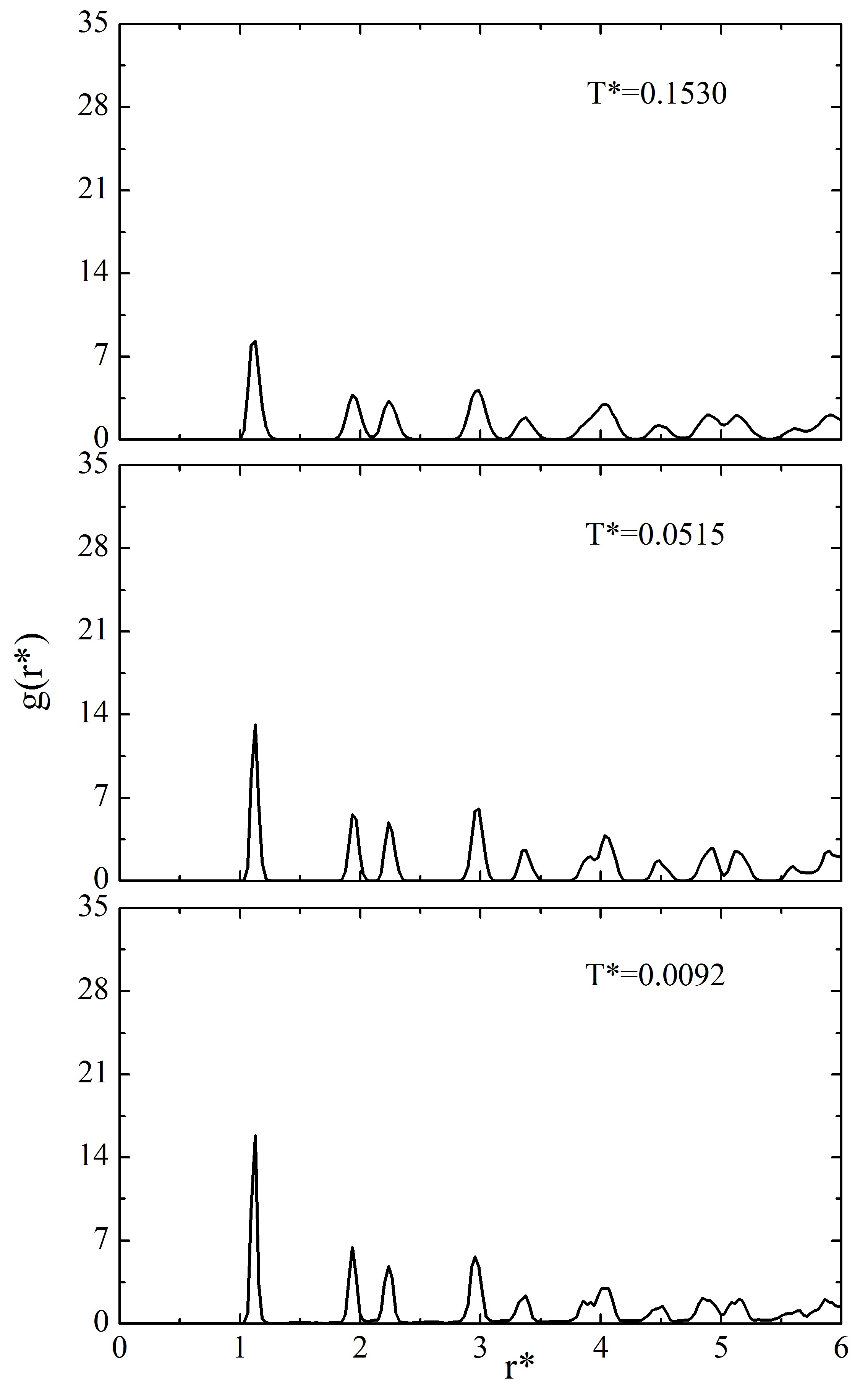}
\caption{Radial Distribution Function, IE parameters and  $\rho$*= 0.907, higher than
the corresponding to liquid silicon density.}
\label{figrdfie35}
\end{figure}

\begin{figure}[htpb]
\centering
\includegraphics[width=6cm,height=9.6cm]{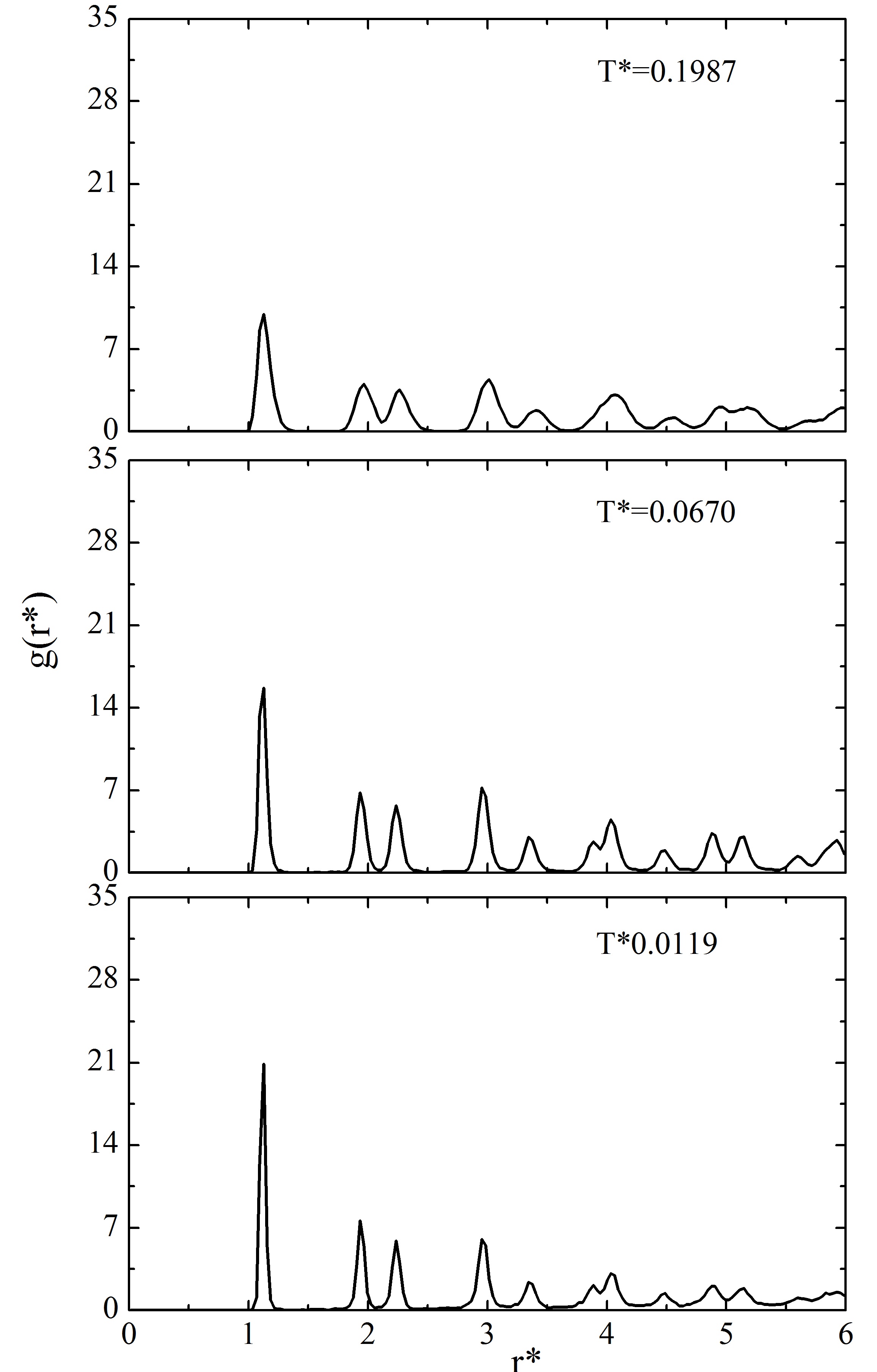}
\caption{Radial Distribution Function, SW parameters and  $\rho$*= 0.690, higher than
the corresponding to liquid silicon density.}
\label{figrdfsw35}
\end{figure}

At $\rho^*_{IE}$ ($\rho^*_{SW}$)= 0.907 (0.690), we obtained the RDF's shown in Figs \ref{figrdfie35} and \ref{figrdfsw35}. For the SW case, the RDF have similar behavior as described above. We can conclude that there is a negligible change with increasing density.

However, for the IE case, the last peaks are almost lost when the density increases; this loss occurs roughly around r*= 3 for T*= 0.0092 and about  r*= 4 for T*=0.0515, being this behavior more noticiable at T*= 0.0092. At T*= 0.1530, the RDF does not show a diferent behavior to that presented at $\rho$*= 0.604.

\begin{figure}[!hptb]
\centering
\includegraphics[width=6cm,height=8cm]{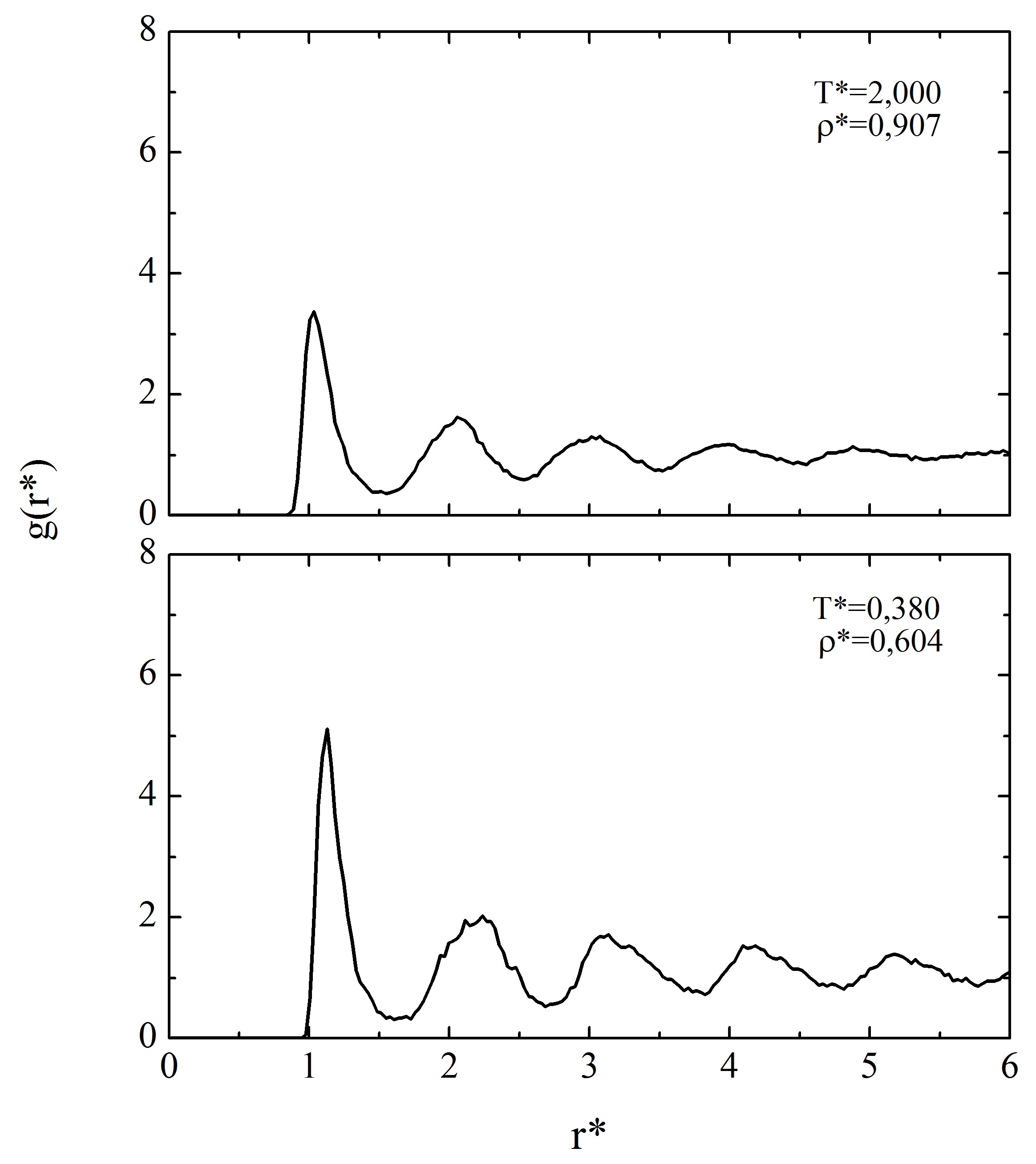}
\caption{High temperature radial distribution function calculated with IE parameters. Density $\rho*$ and temperatures as indicated. A typical shape for a liquid appears.}
\label{figrdfief}
\end{figure}

\begin{figure}[!hptb]
\centering
\includegraphics[width=6cm,height=8cm]{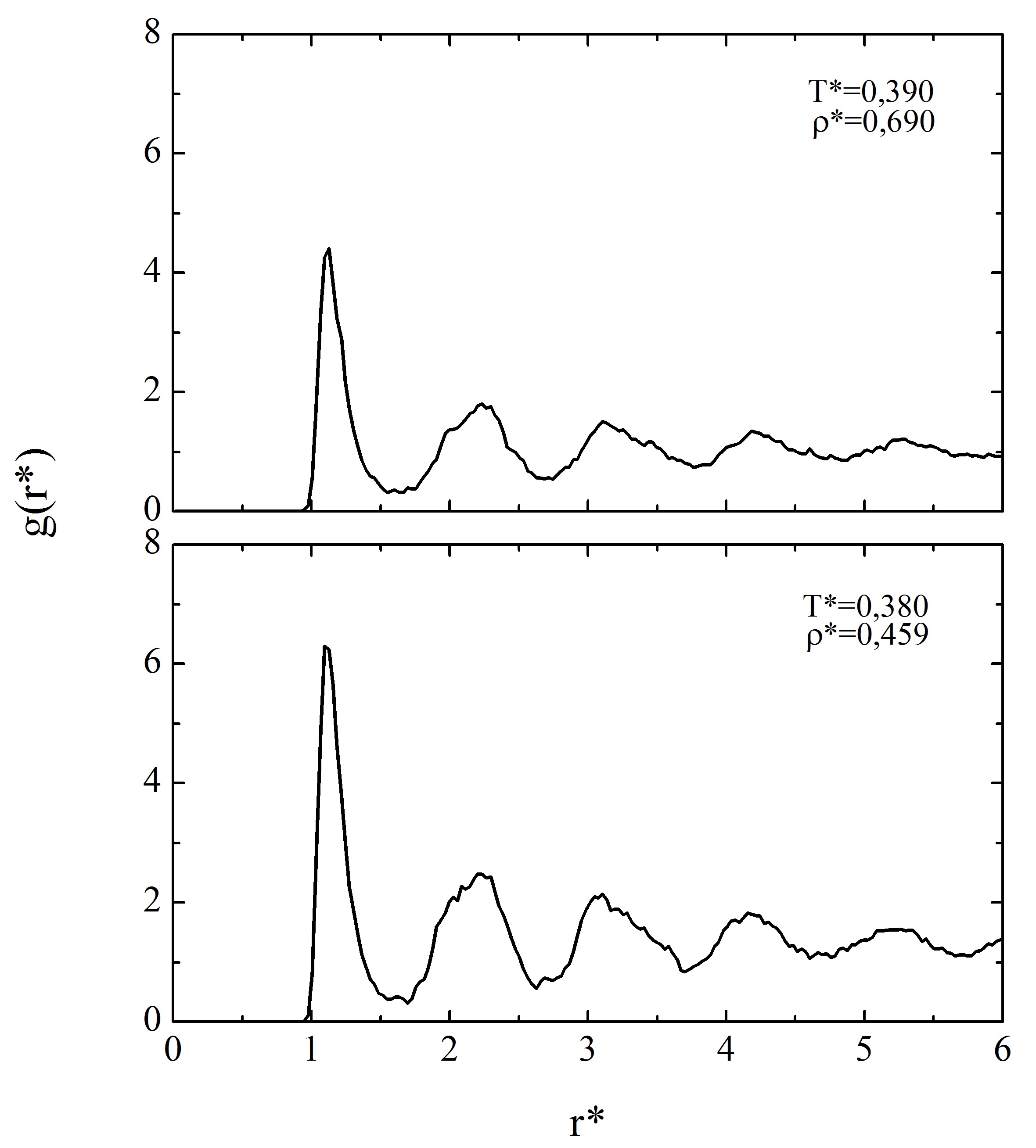}
\caption{High temperature radial distribution function calculated with SW parameters. Density $\rho*$ and temperatures as indicated. A typical shape for a liquid appears.}
\label{figrdfswf}
\end{figure}

Finally, Figs. \ref{figrdfief} and \ref{figrdfswf} show RDF's at very high temperatures, and clearly their shapes correspond to the liquid phase, indicating that melting ocurrs at high temperature. These high temperatures may be due to the restriction on atoms to remain in two-dimensional motion, so it takes more energy to break \emph{sp}$^2$ bonds, which are much less common for Si \cite{Lebegue-2009}.

\section{Conclusions}

We have performed 2D molecular dynamics simulations to study the radial distribution function of silicene.
A Lennard-Jones potential was considered and two sets of parameters were used. 
The RDF's obtained correspond to a two-dimensional close-paked hexagonal lattice, i. e. the plane (111) of the FCC structure.
No significant differences between results obtained with the two sets of parameters were found. Our results suggest a very high melting temperature for the
silicene. More detailed studies with more appropiate potentials for covalent semiconductors are in progress.

\newpage
LMM thanks L. C. Lew Yan Voon for introducing her to the topic.

\begin{acknowledgments}
Partially supported by CONACYT, Grant. CB-2009-133251 and VIEP-BUAP.
\end{acknowledgments}


\newpage
\begin{thebibliography}{99}

\bibitem{Kroto-1985}H. W. Kroto, J. R. Heath, S. C. Obrien, R. F. Curl and R. E. Smalley, Nature {\bf 318}, 162 (1985).

\bibitem{Iijima-1993}S. Iijima and T. Ichihashi, Nature {\bf 363}, 603 (1993).

\bibitem{Novoselov-2004}K. S. Novoselov, A. K. Geim, S. V. Morozov, D. Jiang, Y. Zhang, S. V. Dubonos, I. V. Grigorieva and A. A. Firsov, Science {\bf 306}, 666 (2004).

\bibitem{Castro-2009} A. H. Castro-Neto, F. Guinea, N. M. R. Peres,  K. S. Novoselov and A. K. Geim, Rev. Mod. Phys. {\bf 81}, 109 (2009).

\bibitem{DasSarma-2011}S. Das Sarma, Shaffique Adam, E. H. Hwang and Enrico Rossi, Rev. Mod. Phys. {\bf 83}, 407 (2011).

\bibitem{Aufray-2010}B. Aufray, A. Kara, S. Vizzini, H. Oughaddou, C. Léandri, B.
 Ealet and G. Le Lay, Appl. Phys. Lett. {\bf 96}, 183102 (2010).

\bibitem{Cahangirov-2009}S. Cahangirov, M. Topsakal, E. Akturk, H. Sahin and S. Ciraci, Phys. Rev. Lett. {\bf 102}, 236804 (2009).

\bibitem{Houssa-2010}M. Houssa, G. Pourtois, V. V. Afanas'ev, and A. Stesmans, Appl. Phys. Lett. {\bf 97}, 112106 (2010).

\bibitem{Yu-2010}S. Yu. Davydov, Physics of the Solid State {\bf 52}, 184 (2010).

\bibitem{Takeda-1994}K. Takeda and K. Shiraish, Phys. Rev. {\bf  B 50}, 14916 (1994). G. G. Guzman-Verri and L. C. Lew Yan Voon, Phys. Rev. {\bf  B 76}, 075131 (2007).

\bibitem{De Padova-2008}P. De Padova, C. Léandri, S. Vizzini, C. Quaresima, P. Perfetti, B. Olivieri, H. Oughaddou, B. Aufray and G. Le Lay , Nano Lett. {\bf 8}, 2299 (2008).

\bibitem{Lebegue-2009}S. Leb\'egue and O. Eriksson, Phys. Rev. {\bf  B 79}, 115409 (2009).

\bibitem{Lalmi-2010}B. Lalmi, H. Oughaddou, H. Enriquez, A. Kara, S. Vizzini, B. Ealet and B. Aufray, Appl. Phys. Lett. {\bf 97}, 223109 (2010).

\bibitem{Fleurence-2011}A. Fleurence, R. Friedlein, Y. Wang and Y. Yamada-Takamura, \emph{Experimental evidence for silicene on $ZrB_2$ (0001)}, Symposium on Surface and Nano Science 2011 (SSNS'11), Shizukuishi: Japan (2011).


\bibitem{Boon-2007}Boon K. Teo and X. H. Sun, Chem. Rev. {\bf 107}, 1454 (2007).

 \bibitem{Zhang-2002}R.Q. Zhang, S.T. Lee, Chi-Kin Law, Wai-Kee Li, Boon K. Teo, Chem. Phys. Lett. {\bf 364}, 251 (2002). 

\bibitem{Fagan-2001}S.B. Fagan, R. Mota, R.J. Baierle, G. Paiva, A.J.R. da Silva, A. Fazzio, J. Mol. Struct. (Theochem) {\bf 539}, 101 (2001).

\bibitem{Jose-2011}D. Jose and  A. Datta, Phys. Chem. Chem. Phys. {\bf 13}, 7304 (2011).

\bibitem{Ince-2011}A. Ince and S. Erkoc, Comp. Mater. Sci. {\bf 50}, 865 (2011).

\bibitem{Stillinger-1985}F. H. Stillinger and T. A. Weber, Phys. Rev. {\bfseries B 31}, 5262 (1985).

\bibitem{Galashev-2007}A. E. Galashev, V. A. Polukhin, I. A. Izmodenov and O. R. Rakhmanova, Glass Physics and Chemistry {\bf 33} (1), 86 (2007).

\bibitem{nagbody}Web-Site:\url{http://www.astro.inin.mx/mar/nagbody}.

\bibitem{Rodriguez-2011}M.A. Rodr\'iguez-Meza, J. Terrones-Portas y A.K.L. Silva-Montes. Rev. Mex. Fis. Submmitted (2012).
\bibitem{Rapaport-1995} D. C. Rapaport, \emph{The Art of Molecular Dynamics Simulation}, Cambridge University Press, New York, 1995.

\bibitem{Ioffe}Ioffe Physical Technical Institute, \emph{New Semiconductor Materials: Characteristics and Properties}: {\bf Physical Properties of Semiconductors}, Ioffe Institute 1998-2001, (14 Apr. 2011). Web-Site: \url{http://www.ioffe.ru/SVA/NSM/Semicond/Si/index.html}.

\bibitem{Zakharchenko-2011}K. V. Zakharchenko, A. Fasolino, J. H. Los and M.I. Katsnelson, J. Phys.: Condens. Matter {\bf 23}, 202202 (2011).

\bibitem{Bechstedt-2003}F. Bechstedt, \emph{Principles of Surface Physics}, Springer-Verlag, Germany, 2003.

\bibitem{Broughton-1983}J. Q. Broughton and G. H. Gilmer, J. Chem. Phys. {\bf 79}, 5105 (1983).

\bibitem{Fasolino-2007}A. Fasolino, J. H. Los and M. I. Katsnelson, Nature Mater. {\bf 6} (11), 858 (2007).

\bibitem{Zhao-2009}H. Zhao, K. Min and N. R. Aluru, Nano Lett. {\bf 9} (8), 3012 (2009).

\end{thebibliography}
\end{document}